\documentclass[10pt]{article}

\usepackage{amsmath}
\usepackage{amssymb}

\usepackage{graphicx}

\usepackage{cite}

\usepackage{color}

\usepackage[labelfont=bf,labelsep=period,justification=raggedright]{caption}

\makeatletter
\renewcommand{\@biblabel}[1]{\quad#1.}
\makeatother

\date{}

\begin{document}

\begin{flushleft}
{\Large{\textbf{Electrodiffusive model for astrocytic and neuronal ion concentration dynamics}
}}
\\
Geir Halnes$^{1,\ast}$,
Ivar {\O}stby$^{1}$,
Klas H. Pettersen$^{2}$,
Stig W. Omholt$^{3}$,
Gaute T. Einevoll$^{1}$
\\

\

\textsuperscript{\bf{1}} Dept. of Mathematical Sciences and Technology, Norwegian University of Life Sciences, {\AA}s, Norway
\\
\textsuperscript{\bf{2}} Centre for Integrative Genetics, Dept. of Mathematical Sciences and Technology, Norwegian University of Life Sciences, {\AA}s, Norway
\\
\textsuperscript{\bf{3}} Centre for Integrative Genetics, Dept. of Animal and Aqucultural Sciences, Norwegian University of Life Sciences, {\AA}s, Norway
\\
\textsuperscript{$\ast$} E-mail: geir.halnes@umb.no
\end{flushleft}

\section*{Abstract}
Electrical neural signalling typically takes place at the time-scale of milliseconds, and is typically modeled using the cable equation. This is a good approximation when ionic concentrations are expected to vary little during the time course of a simulation. During periods of intense neural signalling, however, the local extracellular K\textsuperscript{+}-concentration may increase by several millimolars. Clearance of excess K\textsuperscript{+} likely depends partly on diffusion in the extracellular space, partly on local uptake by- and intracellular transport within astrocytes. The processes that maintain the extracellular environment typically takes place at the time scale of seconds, and cannot be modeled accurately without accounting for the spatiotemporal variations in ion concentrations. This work presented here consists of two main parts: First, we developed a general electrodiffusive formalism for modeling ion concentration dynamics in a one-dimensional geometry, including both an intra- and extracellular domain. The formalism was based on the Nernst-Planck equations. It ensures (i) that the membrane potential and ion concentrations are in consistency, (ii) global particle/charge conservation, and (iii) accounts for diffusion and concentration dependent variations in resistivities. Second, we applied the formalism to a model of astrocytes exchanging ions with the ECS. Through simulations, we identified the key astrocytic mechanisms involved in K\textsuperscript{+} removal from high concentration regions. We found that a local increase in extracellular K\textsuperscript{+} evoked a local depolarization of the astrocyte membrane, which at the same time (i) increased the local astrocytic uptake of K\textsuperscript{+} (by locally inactivating the outward Kir-current), (ii) suppressed extracellular transport of K\textsuperscript{+}, (iii) increased transport of K\textsuperscript{+} within astrocytes, and (iv) facilitated astrocytic relase of K\textsuperscript{+} in regions where the extracellular concentration was low. In summary, these mechanisms seem optimal for shielding the extracellular space from excess K\textsuperscript{+}.


\section*{Introduction}
Electrical neural signalling is typically modeled using the cable equation, where dendrites and axons are represented as one-dimensional, possibly branching, electrical cables, and the transmembrane potential is the key dynamical variable \cite{Hodgkin1952, Rall1977}. With the possible exception of the signalling molecule Ca\textsuperscript{2+} (see e.g., \cite{Destexhe1996b, Halnes2011}), ion concentrations are typically assumed to be constant. The effect of ionic diffusion (due to concentration gradients) on the net electrical currents is neglected in standard cable theory. Also, resistivities (which in reality depend on ion concentrations) are assumed to be constant. These are often good approximations, as concentrations of the main charge carriers (K\textsuperscript{+}, Na\textsuperscript{+} and Cl\textsuperscript{-}) in the extracellular- (ECS) or intracellular space (ICS) typically vary little at the short time-scale relevant for electrical neural activity ($<100 ms$).

Ion concentrations in neural tissue are actively regulated by ion pumps and membrane co-transporters. Regulation of extracellular ion concentrations is one of the key cellular functions of astrocytes \cite{Wang2008}. These regulatory processes typically take place at a longer time-scale ($>1 s$) where spatiotemporal variations of ionic concentrations are likely to occur. Under such circumstances, diffusion is likely to become a viable transport mechanism, and the standard cable-model fails to give accurate predictions \cite{QianNingSejnowski1989}. Previous astrocyte models have therefore been based on the Nernst-Planck equations, which consider ionic movements due to diffusion as well as electrical migration \cite{Gardner-Medwin1983,Chen2000}. It should be noted that electrodiffusion also may be relevant at short time-scales, e.g., in small intracellular volumes, such as presynaptic terminals or dendritic spines, where intracellular ion concentrations may increase by orders of magnitude within a millisecond \cite{Stockbridge1984, QianNingSejnowski1989}.

Electrodiffusive models simulate the dynamics of ionic concentrations ($c_k$) of several ionic species ($k$), as well as the transmembrane potential ($v_M$). A general concern when both $v_M$ and $c_k$ are modeled, is whether the relationship between the two is consistent. Physically, $v_M$ is determined by the total electrical charge on the inside (or outside) of the membrane, which in turn is uniquely determined by the concentrations ($c_k$) of all ionic species that are present there. In order to make specific problems analytically tractable, or speed up simulations, many models allow the strict dependence between $c_k$ and $v_M$ to be violated, e.g., by deriving $v_M$ from a reduced equation that does not include all ionic transports in the system \cite{Gardner-Medwin1983, Nygren1999, Chen2000}. In many situations, such approximations may be warranted. However, if applied to general problems, and in particular in long-term simulations, models that do not ensure an internally consistent $c_k-v_M$ relationship may give erroneous predictions. For example, in some heart cell models, ion concentrations have been reported to drift to unrealistic values in long-term simulations, while $v_M$ maintain realistic values \cite{Varghese1997, Endresen2000, Hund2001}.

Qian and Sejnowski developed a model for electrodiffusion in excitable cells, which in a consistent way derives $v_M$ from the intracellular ionic concentrations \cite{QianNingSejnowski1989}. Like the standard cable model, the electrodiffusive model assumes that transport phenomena are essentially one-dimensional. Unlike the standard cable model, the electrodiffusive model includes the concentration dynamics of all involved ions, the diffusive currents arising from intracellular concentration gradients, and the concentration dependent variation of the intracellular resistivities. However, an important limitation with this model is that it only includes intracellular transports, whereas the ECS is assumed to be isopotential and with constant ion concentrations. In reality, the ECS comprises about 20\% of the total neural tissue volume, while the remaining 80\% is the ICS of various cells (mainly astrocytes and neurons). When a large number of cells participate in simultaneous ion exchange with the ECS, the impact on the ion concentrations in the ICS and ECS may be of the same order of magnitude. For example, it is known that the local extracellular K\textsuperscript{+}-concentration can increase by several millimolars (i.e., more than double) during periods of intense neural activity \cite{Gardner-Medwin1983, Dietzel1989, Øyehaug2012}. Furthermore, clearance of excess K\textsuperscript{+} from high concentration regions likely depends partly on diffusion in the ECS, partly on local uptake via astrocytic K\textsuperscript{+}-uptake mechanisms, and partly by intracellular transport within astrocytes \cite{Gardner-Medwin1983, Chen2000, Østby2009, Øyehaug2012}. An understanding of these processes, requires a general electrodiffusive framework that explicitly includes both the ECS and the ICS.

\begin{figure}[!ht]
\begin{center}
\includegraphics[width=4in]{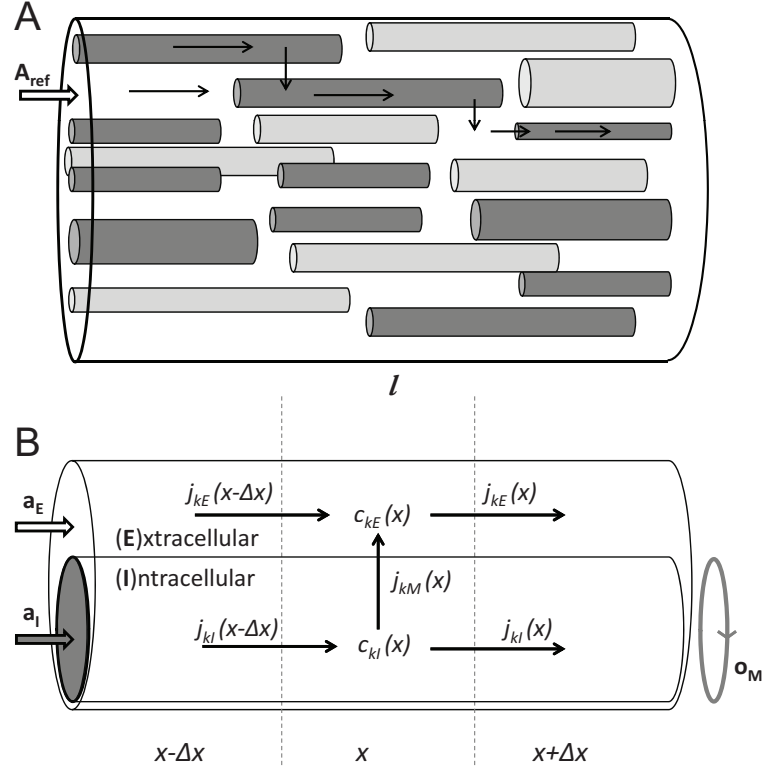}
\end{center}
\caption{
{\bf A two domain-model for ion concentration dynamics in the intra- and extracellular space, when macroscopic transport is essentially one-dimensional.}  (A) A piece of neural tissue with cross section area $A_{ref}$ and an arbitrary extension $l$ in the $x$-direction. The tissue contains cells (dark grey) that participate in the transport process, and cells that do not (light grey). , where $a_I$ is the fraction of $A_{ref}$ that is the ICS of the participatory cell type, $a_E$ is the fraction that is ECS,  (B) The participatory cells represented as an equivalent cylindrical cable ($I$), coated by ECS ($E$). The geometry is specified by three parameters, where $a_I$ and $a_E$ are, respectively, the fractions of $A_{ref}$ occupied by the ICS of participatory cells and the ECS, and $O_M (m^{-1})$ is the amount of membrane area per tissue volume, or, equivalently, the circumference of the equivalent cable. Due to the presence of non-participatory cells, we generally have that $a_I+a_E<1$.
}
\label{Fgeo}
\end{figure}

In situations when macroscopic transport processes are effectively one-dimensional, the complex composition of the tissue (Fig.~\ref{Fgeo}\textit{A}) can be simplified to the two-domain model shown in Fig.~\ref{Fgeo}\textit{B} \cite{Gardner-Medwin1983, Chen2000}. There, the ICS of all cells participating in the transport process have been represented as an equivalent cable ($I$-domain), which is coated by ECS ($E$-domain). The $I$-$E$ system may be pictured phenomenologically as an average single cell coated with the average proportion of available ECS per cell. Such a geometrical simplification was previously motivated for one-dimensional transport phenomena through the glial syncytium \cite{Gardner-Medwin1983, Chen2000}.

In this work, we first derive a simple, general mathematical framework for modeling the dynamics of the membrane potential ($v_M$), the intra- ($c_{kI}$) and extracellular ($c_{kE}$) ion concentrations for a set ($k$) of ionic species, and identify the conditions where the formalism reduces to the standard cable model. Next, we apply the electrodiffusive formalism to model ionic exchange between astrocytes and the ECS, and investigate the relative role of astrocytes in K\textsuperscript{+} removal from high concentration regions. Finally, we provide a discussion of our results, and of the underlying assumptions in our new electrodiffusive framework.

\section*{Results}

\subsection*{Electrodiffusive formalism}
A formalism was derived for computing the ion-concentration dynamics in a geometry as that depicted in Fig.~\ref{Fgeo}\textit{B}. The formalism is summarized in Fig.~\ref{Fbox}, and the derivation is included in the following subsections.

\subsubsection*{Particle conservation}
In Fig.~\ref{Fgeo}\textit{B}, particles in $I$ or $E$ may move along the $x$-axis or across the membrane. In a segment $\Delta x$ of $I$, centered at $x$, and with volume $a_I \Delta x$, the particle concentration dynamics of an ion species $k$ is determined by:

\begin{equation}
-O_M \Delta x j_{kM}(x,t) + a_I j_{kI}(x-\Delta x/2 ,t)) - a_I j_{kI}(x+\Delta x/2,t) = a_I \Delta x \frac{\partial c_{kI}(x,t)}{\partial t},
\label{one}
\end{equation}
\noindent
where the transmembrane- ($j_{kM}$), the intracellular- ($j_{kI}$) and the extracellular ($j_{kE}$) flux densities of particle species $k$, have units mol/(m\textsuperscript{2}s). The first term on the left represents the ionic flux that enter($+$) this segment through the piece of the membrane with area $O_M \Delta x$. The second and third terms represent the ionic fluxes that enter($+$)/leave($-$) the section through the left/right boundaries, with cross section areas $a_I$. If the net flux into the segment is nonzero, the ion concentration will build up over time, according to the right hand side of Eq.~\ref{one}.

We divide Eq.~\ref{one} by $a_I\Delta x$, and take the limit $\Delta x \rightarrow 0$, to obtain the continuity equation on differential form:
\\
\begin{eqnarray}
\frac{\partial j_{kI}(x,t)}{\partial x} + \frac{O_M}{a_I}j_{kM}(x,t) + \frac{\partial c_{kI}(x,t)}{\partial t}=0
\label{ConserveI}
\\
\frac{\partial j_{kE}(x,t)}{\partial x} - \frac{O_M}{a_E}j_{kM}(x,t) + \frac{\partial c_{kE}(x,t)}{\partial t}=0.
\label{ConserveE}
\end{eqnarray}
\noindent
We have also written up the continuity equation for the extracellular domain. By convention, $j_{kM}$ has been defined as positive in the direction from $I$ to $E$.

The axial flux densities are described by the generalized Nernst-Planck equation:
\begin{equation}
j_{kn}(x,t) = -\frac{D_k}{\lambda_n^{2}}\frac{\partial c_{kn}(x,t)}{\partial x}
- \frac{D_k z_k}{\lambda_n^{2} \psi} c_{kn}(x,t) \frac{\partial v_{n}(x,t)}{\partial x},
\label{NP}
\end{equation}
\noindent
where $z_k$ is the valence of ion species $k$, and the index $n$ represents $I$ or $E$. The first term on the right in Eq.~\ref{NP} is the diffusive flux density ($j_{kn}^{d}$), driven by the concentration gradients, and the last term is the field flux density ($j_{kn}^{f}$), i.e., the flux density due to ionic migration in the electrical field. The effective diffusion constant $D_k^* = D_k/\lambda_n^2$ is composed of the diffusion constant $D_k$ in dilute solutions and the tortuosity factor $\lambda_n$, which summarizes the hindrance imposed by the cellular structures \cite{Nicholson1998, Chen2000}. We use $\psi = RT/F (mV)$, where $R= 8.3144621 J/(mol K)$ is the gas constant, $T$ the absolute temperature, and $F=96,4853365 C/mol$ is Faraday's constant. As in \cite{Chen2000}, we have then implicitly assumed that the Einstein relation holds between the effective diffusion constant and effective electrical mobility.

The formalism is general to the form of $j_{kM}$, which may include contribution from multiple membrane mechanisms, such as ion pumps, co-transporters and ion channels. It is sufficient to require that $j_{kM}$ is known at any point in time given the voltage across the membrane, the ionic concentrations on either side of the membrane, and possibly some additional local information ($\tilde{m}_1, \tilde{m}_2, etc.$) reflecting the local state of the membrane:
\begin{equation}
j_{kM}(x,t) = f(c_{kI}(x,t), c_{kE}(x,t), v_M(x,t), \tilde{m}_1(x,t),  \tilde{m}_2(x,t), ...).
\label{membraneflux}
\end{equation}

As boundary conditions, we shall apply the sealed-end condition, i.e., we assume that no fluxes enter or leave through the ends ($x=0$ and $x=l$) of $I$ or $E$:
\begin{equation}
j_{kn}(0,t) = j_{kn}(l,t) = 0.
\label{bc}
\end{equation}
\noindent

Equations~\ref{ConserveI} -~\ref{ConserveE}, together with with Eqs.~\ref{NP},~\ref{membraneflux} and ~\ref{bc}, specify the system we want to solve.
Before we derive the electrodiffusive formalism for this problem, we recall how the standard cable equation can be derived from the principles of particle conservation.

\subsubsection*{Charge conservation}
The particle conservation laws (Eqs.~\ref{ConserveI}-~\ref{ConserveE}) can be transformed to charge conservation laws by the use of the general relations (see e.g., \cite{Koch1999}):
\begin{equation}
\rho_n(x,t) = F\sum_{k}{z_k c_{kn}(x,t)} + \rho_{sn}(x)
\label{rho def}
\end{equation}
\noindent
\begin{equation}
i_M(x,t) = F\sum_{k}{z_k j_{kM}(x,t)}
\label{iM def}
\end{equation}
\begin{equation}
i_n(x,t) = F\sum_{k}{z_k j_{kn}(x,t)}.
\label{in def}
\end{equation}
\noindent
Here, $\rho_n$(C/m\textsuperscript{3}) is the charge density, $i_M$ (A/m\textsuperscript{2}) is the transmembrane current density, and $i_n$ (A/m\textsuperscript{2}) is the axial current density. For practical purposes, we have included a density of static charges ($\rho_{sn}$) in Eq.~\ref{rho def}, representing contributions from ions/charged molecules that are not considered in the conservation equations. If the set $c_{kn}$ include all present species of ions, then $\rho_{sn}=0$. To keep notation compact, we from here on omit the functional arguments $(x,t)$.

If we multiply the particle conservation laws (Eqs.~\ref{ConserveI} -~\ref{ConserveE}) by $F z_k$, take the sum over all ion species, $k$, and use Eqs.~\ref{rho def} -~\ref{in def}, we obtain the equivalent laws for charge conservation:
\begin{eqnarray}
a_I \frac{\partial i_{I}}{\partial x} + O_M i_{M}+ a_I \frac{\partial \rho_I}{\partial t}=0
\label{chargeconservationI}
\\
a_E \frac{\partial i_{E}}{\partial x} - O_M i_{M}+ a_E \frac{\partial \rho_E}{\partial t}=0.
\label{chargeconservationE}
\end{eqnarray}
\noindent
Note that the last term only depends on the mobile ions, as $\partial \rho_{sn}/\partial t = 0$.

\subsubsection*{Standard cable equation}
The standard cable equation may be derived by combining the charge conservation laws (Eqs.~\ref{chargeconservationI} -~\ref{chargeconservationE}) with three simplifying assumptions: (i) $E$ is assumed to be isopotential and with zero resistivity, (ii) the membrane is a parallel-plate capacitor, and (iii) ion concentrations are effectively constant, i.e., diffusive currents are negligible and resistivities (see Eq.~\ref{resistivities} below) are constant.

Assumption (i) implies that we only need to consider charge conservation in $I$ explicitly. To obtain the cable equation in the standard form, we must express $\rho_I$ and $i_{I}$ in Eq.~\ref{chargeconservationI} in terms of $v_M$ and $\partial v_M / \partial x$.

Assumption (ii) allows us substitute $v_M$ for $\rho_I$. A capacitor with capacitance $\delta C$ separates a charge $\delta Q$ from the opposite charge $-\delta Q$, and generates a voltage difference $v = \delta Q/\delta C$. The charge inside a piece ($\delta x$) of membrane with area $O_M \delta x$ is $\delta Q_I = \rho_I a_I \delta x$. The capacitance of this piece of membrane is $\delta C = C_M O_M \delta x$, where $C_M$ denotes the membrane capacitance per membrane area. We therefore obtain:
\noindent
\begin{equation}
v_M = \frac{\delta Q_I}{\delta C} =  \frac{\rho_I a_I \delta x}{C_M O_M \delta x} = \frac{a_I}{O_M} \frac{\rho_I}{C_M}.
\label{VMold}
\end{equation}
\noindent

According to assumption (iii), diffusive currents are negligible, and Eq.~\ref{NP} reduces to:
\begin{equation}
j_{kI} = j_{kI}^{f} =  - \frac{D_k z_k}{\lambda_I^{2} \psi} c_{kI} \frac{\partial v_{I}}{\partial x}.
\label{jf}
\end{equation}
\noindent
If we insert Eq.~\ref{jf} into Eq.~\ref{in def}, we see that the axial current density obeys Ohm's current law:
\begin{equation}
i_{I} = i_{I}^{f} = -\sum_{k}{\frac{F D_k z_k^{2}}{\lambda_I^{2} \psi} c_{kI}}\frac{\partial v_{I}}{\partial x}
=  -\frac{1}{r_I} \frac{\partial v_{I}}{\partial x},
\label{ifield}
\end{equation}
\noindent
where we have identified the resistivity, $r_n (\Omega m)$:
\begin{equation}
\frac{1}{r_n} = \sum_{k}{\frac{F D_k z_k^{2}}{\lambda^{2}_n \psi} c_{kn}},
\label{resistivities}
\end{equation}
\noindent
in the ICS ($n=I$). Finally, we insert Eqs.~\ref{ifield} and~\ref{VMold} into Eq.~\ref{chargeconservationI} to obtain the cable equation:
\begin{equation}
-\frac{a_I}{O_M} \frac{1}{r_I} \frac{\partial^2 v_M}{\partial x^2} + i_{M}+ C_M \frac{\partial v_{M}}{\partial t}= 0.
\label{standardcable}
\end{equation}
Note that $r_n$ generally depends on $c_{kn}$. However, we have here assumed that $r_{n}$ is constant (cf. assumption (iii)).  Furthermore, we have used the identity: $\partial v_I/\partial x = \partial v_M/\partial x$, which follows from the definition
\begin{equation}
v_M = v_I-v_E,
\label{v_Mdef}
\end{equation}
\noindent
together with the assumption (i) that $E$ is isopotential. Eq.~\ref{standardcable} is the most commonly used form of the cable equation, although there are versions that also explicitly considers the extracellular domain \cite{Rall1977}.

\subsubsection*{Two-domain electrodiffusive model}
The cable equation only considers the net electrical transports, and "hides" the underlying transports of different ionic species. We now develop the electrodiffusive formalism for computing the ion-concentration dynamics. Like in standard cable theory, we limit the study to the one-dimensional geometry in Fig.~\ref{Fgeo}\textit{B}. Unlike standard cable theory, we explicitly consider both domains $I$ and $E$, and we do not neglect diffusive currents nor concentration dependent variations of the resistivities.

The conservation equations (Eqs.~\ref{ConserveI} -~\ref{ConserveE}), with the Nernst-Planck equation (Eq.~\ref{NP}) for $j_{kn}$ specify the system we want to solve. As in standard cable theory, the formalism is general to the form of $j_{kM}$ (Eq.~\ref{membraneflux}). With $N$ ion species, Eqs.~\ref{ConserveI} -~\ref{ConserveE} represent a system of $2N+3$ variables which are functions of $x$ and $t$. These are the $2N$ concentration variables ($c_{kn}$ for $k = 1,2,... N$ and $n = E,I$), and the three additional variables ($v_M, \partial v_I /\partial x$ and $\partial v_E /\partial x$) occurring in the expressions for the flux densities.

To reduce the number of independent variables to the $2N$ state variables ($c_{kn}$) we need three conditions relating $v_M$, $\partial v_I /\partial x$ and $\partial v_E /\partial x$ to $c_{kn}$. The first two conditions we recognize from standard cable theory, namely (C1) that $v_M$ is determined by the charge density (Eq.~\ref{VMold}), and (C2) the general definition of (Eq.~\ref{v_Mdef}) of $v_M$. In the two-domain case, we use the additional condition (c3):
\begin{equation}
a_I \rho_I = - a_E \rho_E.
\label{CS}
\end{equation}
\noindent
We shall refer to C3 as the \textit{charge symmetry condition}. Its origin is explained below.

According to condition C1, $v_M$ is given by Eq.~\ref{VMold}:
\begin{equation}
v_M = \frac{a_I}{C_M O_M} \rho_I = \frac{a_I}{C_M O_M} (F\sum_{k}{z_k c_{kI}} + \rho_{sI})
\label{VMI}
\end{equation}
\noindent
where we have inserted Eq.~\ref{rho def} for $\rho_I$, so that $v_M$ is expressed in terms of ionic concentrations. Equivalently, we may also express $v_M$ in terms of the ion concentrations in the ECS:
\begin{equation}
v_M = - \frac{a_E}{C_M O_M} \rho_E = - \frac{a_E}{C_M O_M} (F\sum_{k}{z_k c_{kE}} + \rho_{sE}),
\label{VME}
\end{equation}
where the negative sign follows from the convention that $v_M$ is positive when $I$ is positively charged. By demanding consistency between Eq.~\ref{VMI} and Eq.~\ref{VME}, we can derive Eq.~\ref{CS}, which is the \emph{charge symmetry condition} (C3). It implies that the charge on the inside of a piece of membrane is equal in magnitude and opposite in sign to the charge on the outside. C1 and C3 are both implicit when the membrane is assumed to be a parallel plate capacitor. C3 is also related to the issue of electroneutrality (see Discussion).

The next step is to express the voltage gradients ($\partial v_n /\partial x$) in terms of ionic concentrations. The constraints C2 (Eq.~\ref{v_Mdef}) and C3 (Eq.~\ref{CS}) allow us to derive two independent equations that relate $\partial v_E /\partial x$ and $\partial v_I /\partial x$. The first equation is obtained by differentiating Eq.~\ref{v_Mdef}:
\begin{equation}
\frac{\partial v_M(x)}{\partial x} = \frac{\partial v_I(x)}{\partial x}-\frac{\partial v_E(x)}{\partial x}.
\label{vmgrad}
\end{equation}
\noindent
We recall that $v_M$ is already a known function of ion concentrations (Eq.~\ref{VMI} or Eq.~\ref{VME}).

A second equation relating $\partial v_I /\partial x$ to $\partial v_E /\partial x$ may be derived by combining Eq.~\ref{CS} with the charge conservation laws. If we sum Eqs.~\ref{chargeconservationI} and~\ref{chargeconservationE}, we immediately see that the terms involving $i_M$ cancel out. Due to Eq.~\ref{CS}, also the last terms on the left cancel, so that we are left with:
\begin{equation}
a_I \frac{\partial i_{I}}{\partial x} = - a_E \frac{\partial i_{E}}{\partial x}.
\label{CSdyn0}
\end{equation}
\noindent
\noindent
Due to sealed end-condition (Eq.~\ref{bc}), $i_n(0)=$0, so that Eq.~\ref{CSdyn0} takes the simple form:
\begin{equation}
a_I i_{I} = - a_E i_{E}.
\label{CSdyn}
\end{equation}
\noindent
If the charge symmetry condition (C3) is satisfied at a given time $t=0$ (and we must specify the initial concentrations so that this is true), Eq.~\ref{CSdyn} is the condition that it is satisfied at all times $t$.

We now decompose the current density in a diffusive term and a field term: $i_n = i_{n}^{d}+i_{n}^{f}$, and express $i_n^f$ in terms of Ohm's law (cf. Eq.~\ref{ifield}). If we insert this into Eq.~\ref{CSdyn}, we obtain the second equation relating $\partial v_E /\partial x$ and $\partial v_I /\partial x$:
\begin{equation}
a_I \left(i_{I}^{d}+ \frac{1}{r_I}\frac{\partial v_I}{\partial x}\right) = - a_E \left(i_{E}^{d}+\frac{1}{r_E} \frac{\partial v_E}{\partial x}\right).
\label{CSdyn3}
\end{equation}

\begin{figure}[!ht]
\begin{center}
\includegraphics[width=4in]{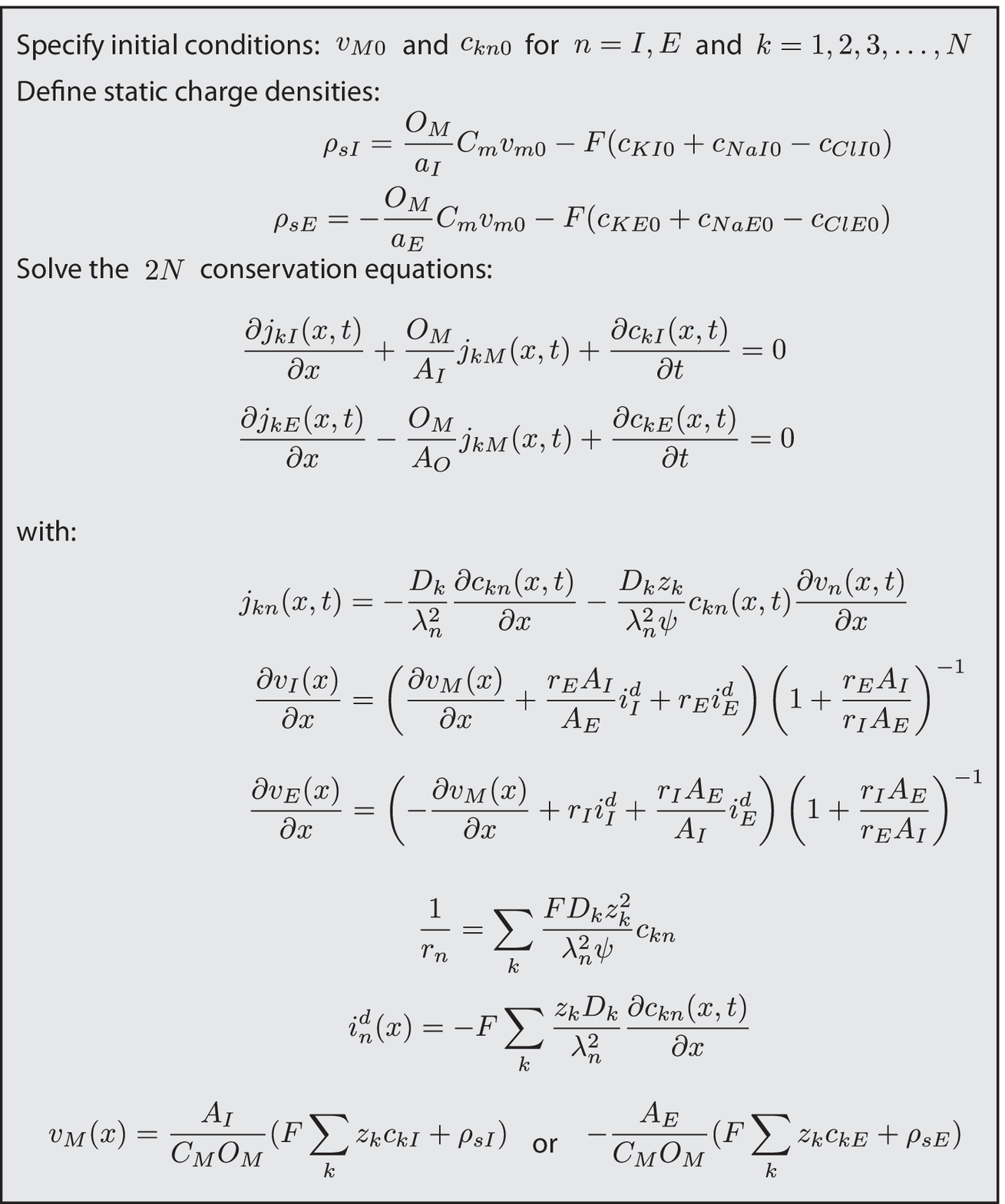}
\end{center}
\caption{
{\bf The two-domain electrodiffusive formalism.} The set of equations summarizes (and fully specify) the electrodiffusive formalism. It is applicable to general problems. The transmembrane currents ($j_{kM}$) need to be specified for any membrane mechanism included in a model.
}
\label{Fbox}
\end{figure}

Finally, Eq.~\ref{vmgrad} and Eq.~\ref{CSdyn3} can be solved for the voltage gradients. After some simple algebra we obtain:
\begin{eqnarray}
\frac{\partial v_I}{\partial x} =
\left(\frac{\partial v_M}{\partial x}+\frac{r_E a_I}{a_E}i_I^{d}+r_E i_E^{d}\right)\left(1+\frac{r_E a_I}{r_I a_E}\right)^{-1}
\label{VIgrad}
\\
\frac{\partial v_E}{\partial x} =
\left(-\frac{\partial v_M}{\partial x}+r_I i_I^{d}+ \frac{r_I a_E}{a_I}i_E^{d}\right)\left(1+\frac{r_I a_E}{r_E a_I}\right)^{-1}.
\label{VEgrad}
\end{eqnarray}
\noindent
Here, $r_n$ is given by Eq.~\ref{resistivities}, $i_n^d$ by Eq.~\ref{NP}, and $v_M$ by Eq.~\ref{VMI} or Eq.~\ref{VME}. All voltage terms are thereby expressed in terms of ionic concentrations. With this, the conservation equations (Eqs.~\ref{ConserveI} -~\ref{ConserveE}) are fully specified, and can be solved numerically with appropriate boundary conditions. The final set of equations is summarized in Fig.~\ref{Fbox}.

\subsubsection*{Electrodiffusive formalism vs cable equation}
From Eq.~\ref{chargeconservationI} for charge conservation in $I$, we may derive a differential equation for the time development of $v_M$. We use Eq.~\ref{VMI} to substitute $v_M$ for $\rho_I$. Furthermore, we use the decomposition $i_I = i_I^d + i_I^f$, with Eq.~\ref{ifield} for $i_I^f$, and Eq.~\ref{VIgrad} for $\frac{\partial v_I(x)}{\partial x}$. We then obtain:
\begin{equation}
\frac{a_I}{O_M} \frac{\partial}{\partial x} \left[
\left(i_I^d - \frac{r_E}{r_I} i_E^d - \frac{1}{r_I} \frac{\partial v_M}{\partial x} \right)
\left(1+\frac{r_E a_I}{r_I a_E} \right)^{-1}
\right]
+ i_M + C_M \frac{\partial v_M}{\partial t} = 0.
\label{ourcable}
\end{equation}
\noindent
This is the equivalent to the standard cable equation (Eq.~\ref{standardcable}), for the electrodiffusive two-domain system.

A few notes: Firstly, a corresponding dynamical equation for $v_M$ could have been derived from the extracellular conservation law (Eq.~\ref{chargeconservationE}). Due to the charge symmetry condition, the two equations would be equivalent. Secondly, unlike the standard cable equation, Eq.~\ref{ourcable} does not provide a complete system description, as Eqs.~\ref{ConserveI} -~\ref{ConserveE} must be solved to determine $i_n^d$ and $r_n$. Thirdly, when the ionic concentrations are known, Eq.~\ref{ourcable} is not necessary for computing $v_M$, as $v_M$ can be computed algebraically from Eq.~\ref{VMI}. Eq.~\ref{ourcable} is mainly useful for comparison with the standard cable equation.

We can immediately see that if we make the common assumptions (i) that the extracellular resistivity ($r_E$) is zero, (ii) that the diffusive currents ($i_n^d$) are zero, and (iii) that the intracellular resistivity ($r_I$) is constant, then Eq.~\ref{ourcable} reduces to the standard cable equation (Eq.~\ref{standardcable}). We should note that there are two-domain versions of the cable equation where the first assumption is not made \cite{Rall1977}. The two other assumptions are warranted only in cases when the spatiotemporal variations in ionic concentrations is such that $r_I$ varies little, and $i_n^d\ll i_n^f$ during the time course of a simulation.

\subsection*{Astrocyte Model}
The electrodiffusion-formalism was applied in a one-dimensional model for astrocytes exchanging ions with the ECS. The purpose was to investigate the relative role of astrocytes in K\textsuperscript{+} removal from high concentration regions.

The model was developed for macroscopic transport processes, involving all astrocytes in a piece of tissue. The geometry in Fig.~\ref{Fgeo}\textit{B} was therefore applicable, with $I$ representing a phenomenological "average" astrocyte (the cable, $I$), surrounded by a sheet of ECS (the coating, $E$). The geometrical parameters $a_I$, $a_E$ and $O_M$ have been estimated for astrocytes in neural tissue (see Table 1). For the extension in the $x$-direction, we used $l=300 \mu m$.

\begin{figure}[!ht]
\begin{center}
\includegraphics[width=4in]{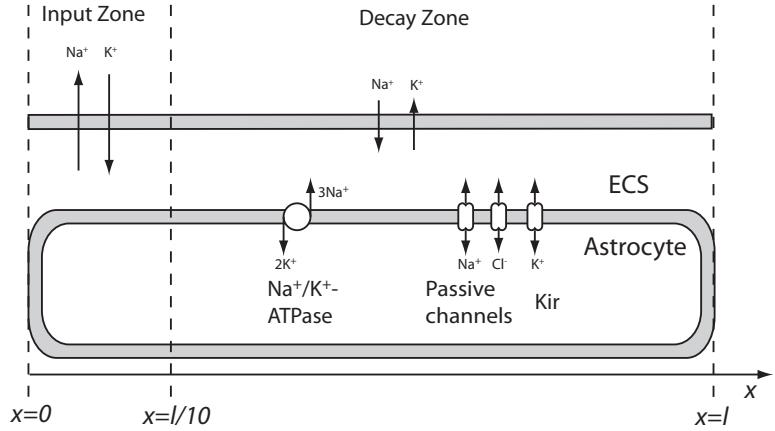}
\end{center}
\caption{
{\bf Astrocyte model.} A representative astrocyte ($I$) exchanging ions with the ECS ($E$) through membrane mechanisms as indicated.
The system input to the ECS was applied in the \textit{input zone}. The output was applied over the whole axis. The \textit{decay zone}
was defined as the part of the axis where no input was applied.
}
\label{Fastro}
\end{figure}

Astrocytic membrane mechanisms were adopted from a previous point-model of an astrocyte \cite{Østby2009}. The included mechanisms were the standard, passive Na\textsuperscript{+} and Cl\textsuperscript{-} channels, the inward rectifying K\textsuperscript{+}-channel (Kir), and the Na\textsuperscript{+}/K\textsuperscript{+}-pump, as sketched in Fig.~\ref{Fastro}. The membrane mechanisms are described in further detail in the \textit{Methods}-section.

We assumed that only the three main charge carriers (K\textsuperscript{+}, Na\textsuperscript{+} and Cl\textsuperscript{-}) contributed to electrodiffusive transport. For the diffusion constants ($D_k$), we used values valid for electrodiffusion in diluted media \cite{Grodzinsky2011}, modified with the tortuosities ($\lambda_n$) estimated in \cite{Chen2000}. The same values have also been used in earlier, related studies \cite{QianNingSejnowski1989, Chung1999}. All relevant model parameters are listed in Table 1.

As external input to the system (mimicking the effect of enhanced, local neuronal activity), we used a constant K\textsuperscript{+}/Na\textsuperscript{+}-exchange (adding K\textsuperscript{+}, subtracting Na\textsuperscript{+}) in a selected region ($0<x<l/10$) of the ECS between $t=100 s$ and $t=400 s$:
\begin{equation}
\begin{array}{lc}
j_K^{in} = - j_{Na}^{in} = j_{in} & \mbox{for $0<x<l/10$ \&  $100<t<400$}\\.
\end{array}
\label{input}
\end{equation}
Conversely, the model output was a \textit{concentration-dependent} Na\textsuperscript{+}/K\textsuperscript{+}-exchange (subtracting K\textsuperscript{+}, adding Na\textsuperscript{+}), occurring over the entire astrocyte axis, causing the extracellular K\textsuperscript{+}-concentration to decay towards the resting concentration. The output could represent the uptake of K\textsuperscript{+}/release of Na\textsuperscript{+} by other neurons via K\textsuperscript{+}/Na\textsuperscript{+}-exchangers and other uptake mechanisms:
\begin{equation}
\begin{array}{lc}
j_K^{out} = - j_{Na}^{out} = -  k_{dec}(c_K-c_{K0}) & \mbox{for all $x$ \&  $t$}\\
\label{output}
\end{array}
\end{equation}
The decay factor ($k_{dec}$) was set to a realistic value for maximal neuronal K\textsuperscript{+}/Na\textsuperscript{+}-exchange under physiological conditions (see Table 1). The input flux density ($j_{in}$) was specified to a value that gave a steady-state K\textsuperscript{+}-concentration of about $10 mM$ in the input region ($0<x<l/10$) during constant input condition (see Fig.~\ref{Fconc}).

\begin{table}[!ht]
\caption{\bf{Model parameters}}
\begin{tabular}{|l|r|l|}
  \textbf{Parameter} & \textbf{Value} & \textbf{Reference} \\
  $l$ (length of astrocyte) & $300 \mu m$ &  \\
  $a_I$ (astrocyte volume/total tissue volume) & 0.4 & \cite{Chen2000} \\
  $a_E$ (ECS volume/total tissue volume) & 0.2 & \cite{Chen2000} \\
  $O_M$ (glia membrane area/total tissue volume) & $4.8 \times 10^{5} m^{-1}$  &  \cite{Chen2000} \\
  $D_K$ (K\textsuperscript{+} diffusion constant) & $1.96\times 10^{-9} m^2/s$ &  \cite{QianNingSejnowski1989, Chung1999, Grodzinsky2011} \\
  $D_{Na}$ (Na\textsuperscript{+} diffusion constant) & $1.33\times 10^{-9} m^2/s$ & \cite{QianNingSejnowski1989, Chung1999, Grodzinsky2011}\\
  $D_{Cl}$ (Cl\textsuperscript{-} diffusion constant) & $2.03\times 10^{-9} m^2/s$  & \cite{QianNingSejnowski1989, Chung1999, Grodzinsky2011} \\
  $\lambda_I$ (intracellular tortuosity) & 3.2 &  \cite{Chen2000} \\
  $\lambda_E$ (intracellular tortuosity) & 1.6  & \cite{Chen2000} \\
  $C_{m}$ (specific membrane capacitance) & $1 \mu F/cm^2$ & \cite{Østby2009}\\
  $g_{K0}$ (baseline K\textsuperscript{+}-conductance)& $16.96 S/m^2$  & \cite{Østby2009}\\
  $g_{Na0}$ (baseline Na\textsuperscript{+}-conductance) & $1 S/m^2$ & \cite{Østby2009} \\
  $g_{Cl0}$ (baseline Cl\textsuperscript{-}-conductance)& $0.5 S/m^2$ & \cite{Østby2009}\\
  $P_{max}$ (maximum Na\textsuperscript{+}/K\textsuperscript{+} pump-rate)& $1.12\times 10^{-6} mol/(Sm^2)$ & \cite{Øyehaug2012} \\
  $K_{KE}$ ($c_{kE}$-threshold for Na\textsuperscript{+}/K\textsuperscript{+} pump)& $1.5 mol/m^3$ & \cite{Øyehaug2012} \\
  $K_{NaI}$ ($c_{NaI}$-threshold for Na\textsuperscript{+}/K\textsuperscript{+} pump) & $10 mol/m^3$ &  \cite{Øyehaug2012} \\
  $c_{KE0}^*$ (initial ECS K\textsuperscript{+}-concentration) & $3.0+0.082 mM$ & \cite{Øyehaug2012} \\
  $c_{KI0}^*$ (initial ECS Na\textsuperscript{+}-concentration) & $100.0-0.041 mM$ & \cite{Øyehaug2012} \\
  $c_{NaE0}^*$ (initial ECS K\textsuperscript{+}-concentration) & $145.0-0.378 mM$ & \cite{Øyehaug2012} \\
  $c_{NaI0}^*$ (initial ICS Na\textsuperscript{+}-concentration) & $15.0+0.189 mM$  & \cite{Øyehaug2012} \\
  $c_{ClE0}^*$ (initial ICS Cl\textsuperscript{-}-concentration)& $134.0-0.29 mM$ & \cite{Øyehaug2012} \\
  $c_{ClI0}^*$ (initial ICS Cl\textsuperscript{-}-concentration) & $5.0+0.145 mM$  & \cite{Øyehaug2012} \\
  $v_{M0}^*$  (initial membrane potential) & $-85+1.4 mV$ & \cite{Øyehaug2012} \\
  $k_{dec}^{\dag}$ (decay factor for $c_{KE}$) & $2.9\times 10^{-8} m/s$ & \cite{Karbowski2009} \\
  $j_{in}$ (constant input in input zone) & $7\times 10^{-8} mol/(m^2s)$ &  \\
  \label{Table}
  \end{tabular}
  \noindent
\begin{flushleft}
\textsuperscript{*} Initial concentrations are given as $c_{kn0} = \mbox{Value from \cite{Øyehaug2012}} + Correction$, where the sum gives rise to a system at rest.\\
\textsuperscript{\dag} The maximum average $Na\textsuperscript{+}/K\textsuperscript{+}$-pump rate for a single neuron was estimated to $A=2\times 10^{-7} mol/(m^2s)$ \cite{Karbowski2009}. We obtained $k_{dec}$ by solving $k_{dec}(c_{KE}^{max} -c_{KE0}) = A$, assuming that $c_{KE}^{max} = 10mM$.
\end{flushleft}
\label{T1}
\end{table}

\subsection*{Ion concentration dynamics in the Astrocyte/ECS system}
\begin{figure}[!ht]
\begin{center}
\includegraphics[width=4in]{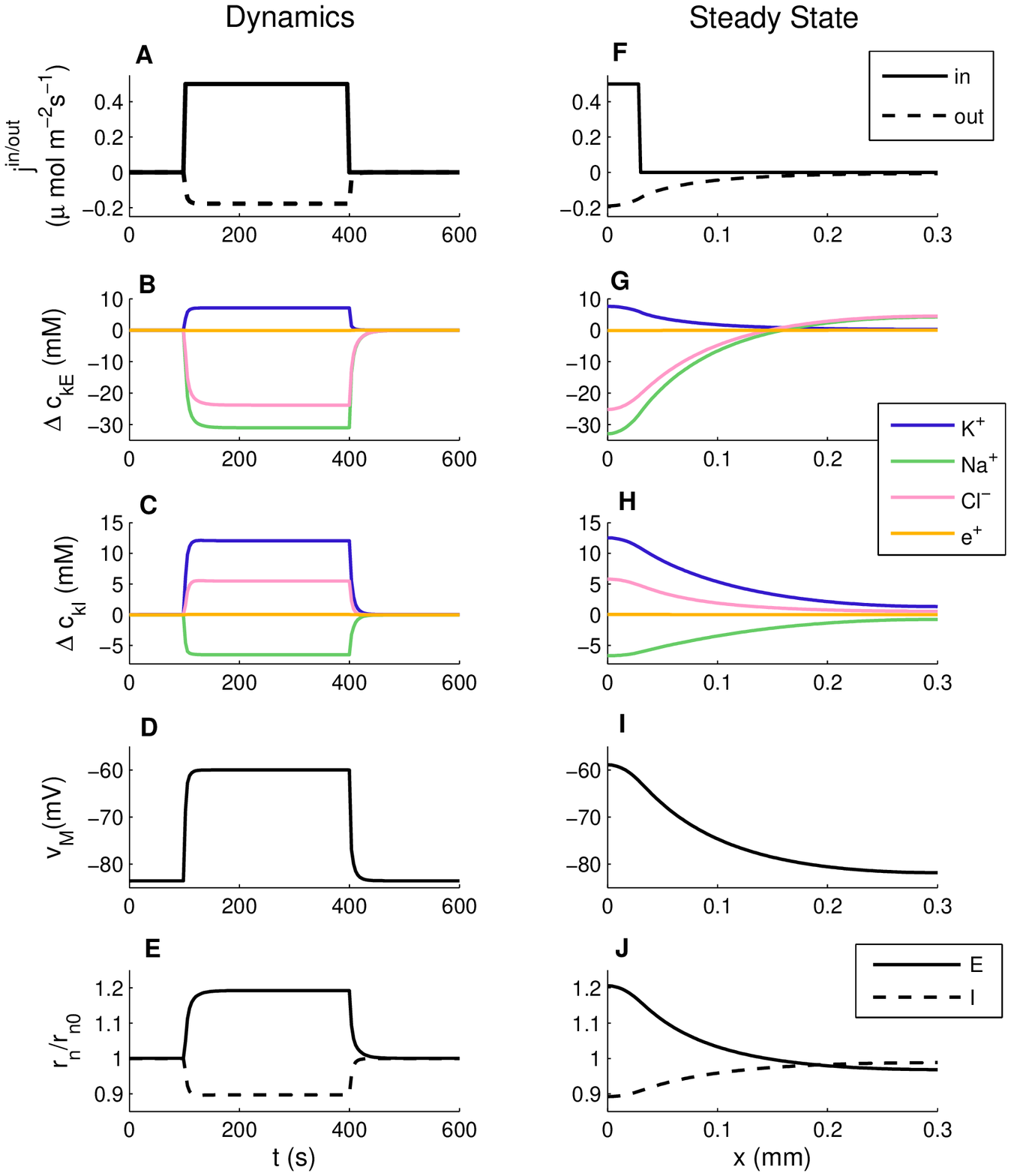}
\end{center}
\caption{
{\bf Dynamics and steady state profiles for the astrocyte/ECS-system.} (A-E) Dynamics of selected variables in a point ($x = l/20 = 0.15 \mu m$)
in the middle of the input zone. The input was given from $t=100 s$ to $t=400 s$.
(F-J) Spatial profiles of selected variables at a time ($t=400 s$) when the system was in steady state.
(B-C, G-H) Ionic concentrations are represented in terms of deviations from resting concentrations: $\Delta c_{kn} = c_{kn}-c_{kn0}$ for $n=I,E$.
(E,J) Resistivities are plotted relative to the resting values $r_{E0} = 1.45 \Omega m$ and $r_{I0}=12.0 \Omega m$.
}
\label{Fconc}
\end{figure}

Fig.~\ref{Fconc}\textit{A-E} shows the dynamics of the atrocyte/ECS system in the middle of the input zone ($x = l/20$). During the input ($100 s<t<400 s$) there was a net influx of K\textsuperscript{+} and a net efflux of Na\textsuperscript{+} to/from the input zone (Fig.~\ref{Fconc}\textit{A}). The constant input caused an increase in $c_K$ and a decrease in $c_{Na}$ both in $I$ and $E$ (Fig.~\ref{Fconc}\textit{B-C}). Although Cl\textsuperscript{-} was not added to the system, $c_{ClI}$ increased on behalf of $c_{ClE}$. The changes in ionic concentrations coincided with a depolarization of the membrane in the input zone (Fig.~\ref{Fconc}\textit{E}), reflecting concentration dependent changes in the reversal potentials of the involved ionic species.

The total output rate (integrated over all $x$) increased with time, due to the general increase in $c_{KE}$. When $c_{KE}$ became sufficiently high, the total output rate coincided with the total input rate, and the system reached steady-state (SS). It took 49 s from the constant input had been turned on until the slowest variable ($|\Delta c_{ClE}|$) reached 99\% of its saturation value. Most variables approached SS significantly faster (e.g., 12 s for $|\Delta c_{KE}|$ and 19 s for $|\Delta v_M|$). When the input was turned off, the system gradually returned to the original resting state.

We here focus on the SS-situation, i.e., on the activity of astrocytes during periods of on-going intense neural activity. Fig.~\ref{Fconc}\textit{F-J} shows the spatial profiles of the input/output, the ionic concentrations and the membrane potential at a time $t_{SS}=400 s$, when the system was in SS. For all variables, deviations from the resting values were greatest in the input zone. There, $c_{KE}$ was about 10 mM during SS, i.e., 7 mM above the resting concentration (3 mM).

The resting potential $v_{M0}=-83.6 mV$ corresponded to concentrations $c_{eI} \simeq 0.18 mM$ and $c_{eE} \simeq 0.36 mM$ of unit charges in the ICS and ECS (cf. Eq.~\ref{ce}). At SS, $v_M$ had increased from the resting potential to about -60 mV, consistent with small absolute changes ($\Delta c_{eI} \simeq 0.05 mM$ and $\Delta c_{eE} \simeq 0.10 mM$) in the concentration of unit charges. As seen in Fig.~\ref{Fconc}\textit{B-C}, these changes were very small compared to the changes $\Delta c_{kn}$ in any of the ionic concentrations. Variations in ion concentrations were thus always so that anions and cations remained closely balanced in numbers, giving rise to a relatively small net charge.

The concentration dependent changes in the resistivities were quite significant. In the input zone, $r_I$ decreased by about 10\%, and $r_E$ increased by about 20\% compared to their basal values (Fig.~\ref{Fconc}\textit{E}).

\subsubsection*{Ion transport pattern in steady state}
During SS, the system output was distributed over the $x$-axis, with equal areas under the input and output curves (Fig.~\ref{Fconc}\textit{F}). In the input zone, the output rate was about 1/3 of the input rate. This means that about 2/3 of the K\textsuperscript{+} that entered the system was transported in the positive $x$-direction, and left the system from the decay-zone (cf. Fig.~\ref{Fastro}).

\begin{figure}[!ht]
\begin{center}
\includegraphics[width=4in]{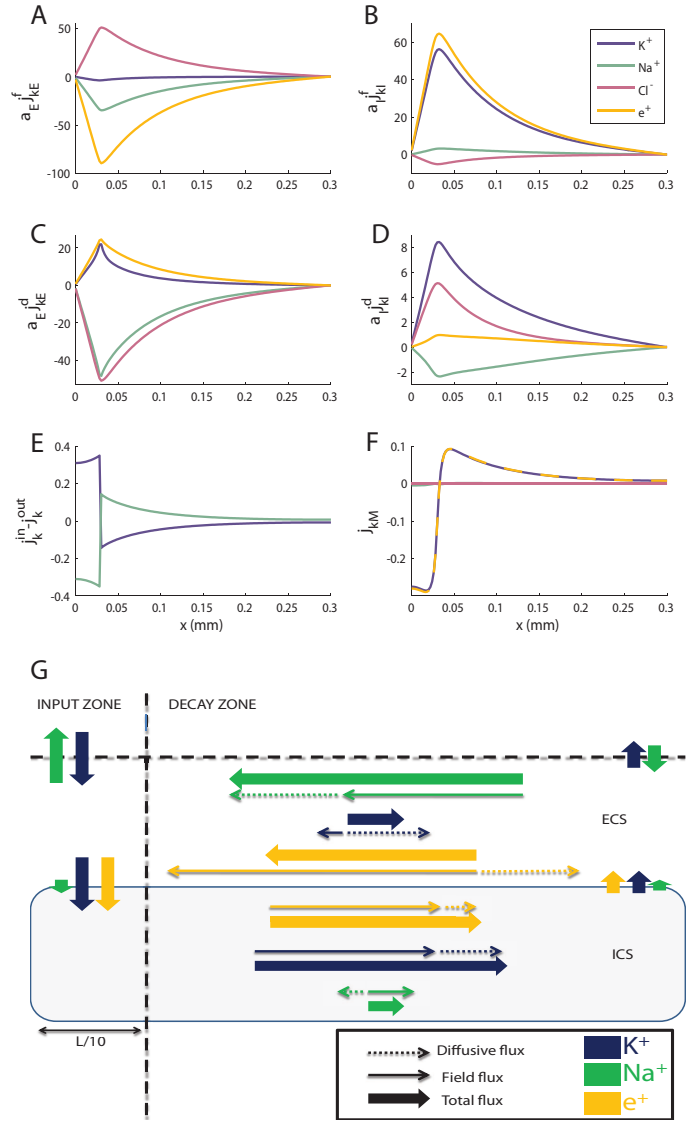}
\end{center}
\caption{
{\bf Transports in the astrocyte/ECS system during steady state ($t=$400s).} (A-B) Flux densities in $E$ and $I$ due to electrical field. (C-D) Flux densities in $E$ and $I$ due to diffusive forces. (A-D) Flux densities $j_{kn}$ were scaled by the relative area fraction $a_n$ so that fluxes in $I$ and $E$ could be compared directly (I.e., $a_Ej_{kE} = a_Ij_{kI}$, gives the same total flux of $k$ in $I$ and $E$). (E) Total flux densities into system ($input-output$). (F) Transmembrane flux densities. (G) Flow chart showing the main transports during steady state. The length of the arrows indicate flux densities, but are not numerically exact.
}
\label{Fflows}
\end{figure}

We wanted to explore the role of the astrocyte in removing K\textsuperscript{+} from the input zone, relative to the role of axial transports in the ECS. To do this, we analyzed the spatial profiles of all ionic flux densities during SS (Fig.~\ref{Fflows}). We distinguished between field flux densities ($j_{kn}^f$) and diffusive flux densities ($j_{kn}^d$). The main transport routes during SS are summarized Fig.~\ref{Fflows}\textit{G}. The net Cl\textsuperscript{-} transport ($j_{ClE}^d+j_{ClE}^f$) was very small, and was not included in the summary.

For Na\textsuperscript{+} and K\textsuperscript{+}, the main transport routes were as follows: Na\textsuperscript{+} entered the system in the decay zone of the ECS, was transported into the input zone, and left the system from the input zone. The main axial transport occurred in the ECS. In contrast, K\textsuperscript{+} entered the system in the input zone, where a major fraction of it crossed the membrane. Transport of K\textsuperscript{+} out from the input zone predominantly took place inside the astrocyte. Outside the input zone (i.e., in the decay zone), the astrocyte released K\textsuperscript{+} to the ECS, from where it eventually left the system. The sum of the Na\textsuperscript{+} and K\textsuperscript{+} transports gave rise to a net electrical current which cycled in the system.

Two basic mechanisms explain the qualitative difference between Na\textsuperscript{+} and K\textsuperscript{+} transports. Both are related to the membrane being most depolarized in the input zone (Fig.~\ref{Fconc}\textit{F}). The first mechanism concerns the transmembrane fluxes. The Na\textsuperscript{+}/K\textsuperscript{+}-exchanger mediated an inward flux of $K^+$ and an outward flux of $Na^+$. The exchanger was counterbalanced by passive fluxes in the opposite direction ($Na^+$ in and $K^+$ out), proportional to $v_M-e_{k}$. In the case of $Na^+$, the passive flux and the exchanger rate were closely balanced across the length of the astrocyte, so that $j_{NaM}$ was small everywhere. In the case of $K^+$, $v_M-e_K$ was very small in the input zone ($e_K \approx -62mV$ (Eq.~\ref{ek}) and $v_M \approx -60mV$), but quite big in the decay zone. In the input zone, $j_{KM}$ was therefore dominated by K\textsuperscript{+}-uptake through the Na\textsuperscript{+}/K\textsuperscript{+}-exchanger, while $K^+$-release through the outward Kir-channel dominated in the decay zone. A similar role of the Kir-channel in spatial K\textsuperscript{+}-buffering has been suggested previously \cite{Kofuji2004}.

A second (and to our knowledge, novel) mechanism for explaining the differences between the Na\textsuperscript{+} and K\textsuperscript{+} transports concerns the axial fluxes. As the astrocyte was most depolarized in the input zone, $I$ had the highest positive charge density and $E$ had the highest negative charge density there. Therefore, the electrical force on Na\textsuperscript{+} and K\textsuperscript{+} (being cations) was in the negative $x$-direction in the ECS ($\partial V_E/\partial x>0$), and in the positive $x$-direction in the ICS ($\partial V_I/\partial x<0$). Transport of K\textsuperscript{+} out from the input zone therefore had the best conditions in $I$, where diffusive and electrical forces were driving the ions in the same direction. For the same reason, Na\textsuperscript{+} transport into the input zone dominated in $E$.

In summary, the local depolarization of the astrocyte induced changes in both the transmembrane and axial flux densities, which both improved the efficiency of the astrocyte in removing K\textsuperscript{+} from the input zone.

\subsubsection*{Consistency of formalism}
In the simulations above, $v_M$ was defined in terms of the charge density in $I$, and computed algebraically by solving Eq.~\ref{VMI} at each time step. Identical results (down to a very small numerical error) were obtained when $v_M$ was defined by the charge density in $E$ (Eq.~\ref{VME}), and when $v_M$ was computed differentially by using Eq.~\ref{ourcable} (results not shown). As all transports are included in Eq.~\ref{ourcable}, the algebraic and differential methods yielded consistent results.

When the input was turned on and off, a small numerical error was introduced in the conservation of ionic concentrations, inducing a small error in the total charge in the system. The relative deviation from global charge neutrality ($Q_{tot} = 0$), defined as $\epsilon_{Qtot} = Q_{tot}/(|Q_{E}|+|Q_{I}|)$, where $Q_I$ and $Q_E$ refer to the total charge in $I$ and $E$, was about $10^{-10}$. This gave rise to a relative deviation from perfect charge symmetry (cf. Eq.~\ref{CS}), defined as $\epsilon_{\rho} = (a_I\rho_{I} - a_E\rho_{E}) /(|a_I\rho_{I}|+|a_E\rho_{E}|)$, which was also on the order of $10^{-10}$ (for all $x$). Accordingly, $v_M$ computed from the charge density in $E$ deviated by a relative factor $\sim 10^{-10}$ from $v_M$ computed from the charge density in $I$. This corresponded to an absolute difference of $\sim 10^{-8}mV$. Errors tended to be somewhat larger when the differential method was used. Then $v_M$ deviated locally by up to $\sim 10^{-6}mV$ from $v_M$ derived from the charge density in $I$ or $E$.

Errors will generally depend strongly on the algorithm used for solving the differential equations, the time step, and the number of compartments in the simulated system. The errors could likely be reduced by using a smoother input signal than the step function in Eq.~\ref{input}. We did not engage in further analysis of the origin of the errors, but were content with their smallness.

\section*{Discussion}
We presented a one-dimensional, electrodiffusive framework for modeling the dynamics of the membrane potential ($v_M$) and the ion concentrations $(c_{kn})$ of all included ion species $(k)$ in an intra- and extracellular domain (Fig.~\ref{Fbox}). The framework could have a broad range of applications within the field of computational neuroscience. In the current work, it was applied to simulate the role of astrocytes in K\textsuperscript{+}-removal from high concentration regions.

\subsection*{Spatial K\textsuperscript{+}-buffering by astrocytes}
The astrocyte/ECS-model provided a mechanistic understanding of how astrocytes may remove K\textsuperscript{+} from high-concentration regions. In summary, the model astrocyte responded to a local extracellular increase in $c_K$ by a local depolarization of the membrane. At the same time, this depolarization (i) increased astrocytic K\textsuperscript{+} uptake in the input zone, (ii) increased astrocytic K\textsuperscript{+}-release outside the input zone, (iii) decreased axial K\textsuperscript{+} transport in the ECS, and (iv) increased axial K\textsuperscript{+} transport inside the astrocyte.

Above, (i-ii) directly concern well documented astrocytic membrane mechanisms.
Increased uptake in high concentration regions (i.e., the input zone in Fig.~\ref{Fastro}) was effectively achieved by inactivation of the outward Kir-channel, facilitating a high net uptake through the K\textsuperscript{+}/Na\textsuperscript{+}-exchanger. This is in agreement with previous experimental findings \cite{Wang2008, Bay2012}. In addition  (iii-iv), we also found that the astrocyte induced changes in the intra- and extracellular voltage gradients that facilitated intracellular K\textsuperscript{+}-transport. Regulation of the longitudinal transport represents a (to our knowledge) novel mechanism that astrocytes may utilize to shield the extracellular space from excess K\textsuperscript{+}. All these effects (i-iv) taken together turned the astrocyte into an efficient sluice for removing K\textsuperscript{+} from the input zone.

The input to the astrocyte/ECS model was a K\textsuperscript{+}/Na\textsuperscript{+} exchange in the ECS. Within a few seconds, this evoked large concentration gradients in the ECS and ICS. These circumstances differ from the constant-background conditions assumed in most neural models, but are physiologically realistic under periods of intense neural activity \cite{Dietzel1989, Chen2000}. As we saw in Fig.~\ref{Fconc}, ionic concentrations changed by several mM before the system reached SS, while $c_{eE}$ and $c_{eI}$ remained bounded by the extreme electrical forces that would be associated with high absolute values of $v_M$. When the system reached SS, the resistivity had increased by up to $\sim 20\%$ from the resting value, and the diffusive current in the ECS was about 25-30\% of the field current. Hence, the predictions made by the electrodiffusive model differed significantly from what would be predicted by standard cable theory, where diffusive currents and concentration dependent variations in resistivities are neglected.

Astrocytes are known to possess several membrane mechanisms that were not included in the current model. K\textsuperscript{+}-uptake by $Na-K-2Cl$-cotransporters and $K-Cl$-cotransporters \cite{Østby2009} are two candidate mechanisms that could affect the simulated results.

\subsection*{Macroscopic transports vs. single cell models}
The astrocyte/ECS-model was represented phenomenologically as a single astrocyte coated with the average proportion of available ECS per astrocyte (Fig.~\ref{Fastro}). This geometrical representation is motivated for macroscopic transport processes, when a large number of astrocytes perform the same function simultaneously \cite{Chen2000}. For spatial K\textsuperscript{+}-buffering, this was a reasonable assumption, as the input was a change in the ion-concentrations in the ECS, shared by all present astrocytes.

If we instead wanted to study a cell specific signal, such as the response of a single astrocyte to a transmembrane current injection, the geometrical representation in Fig.~\ref{Fgeo}\textit{B} would be less appropriate. Firstly, the notion of the ECS as a relatively thin coating following a single cell is only motivated at the macroscopic "average transport"-level. Secondly, $a_I$ and $O_M$ would refer to the single participatory astrocyte, and not all astrocytes intersected by $A_{ref}$. Then we would expect that $a_I<<a_E$, as a single active cell would have a significantly larger proportion of the ECS to its own disposal. In single-cell models it is common to assume that conditions in $E$ are constant, so that only $I$ is modeled explicitly. In this limit, the electrodiffusive formalism reduces to the one-domain model presented previously by Qian and Sejnowski \cite{QianNingSejnowski1989}.

\subsection*{Relationship to other electrodiffusive modeling schemes}
The framework presented here is essentially an expansion of the one-domain model by Qian and Sejnowski \cite{QianNingSejnowski1989} to a two-domain model that includes both the ECS and ICS. Like the one-domain model, the framework ensures (i) a consistent relationship between $v_M$ and $c_k$. Unlike the one-domain model, the framework ensures (ii) global particle/charge conservation, and (iii) that the charges on either side of a piece of membrane must be equal in magnitude and opposite in sign ($\delta Q_I = -\delta Q_E$). The latter constraint is implicit when the the membrane is assumed to be a parallel plate capacitor, an assumption made in most models of excitable cells (see e.g., \cite{Hodgkin1952, Rall1977, Koch1999, QianNingSejnowski1989}). It is also related to the topic of electroneutrality.

Electroneutrality in electrodiffusive models of biological tissue has been the topic of many discussions \cite{Agin1967, Leonetti1998a, Feldberg2000}. It is relevant for how the electrical potential ($v$), occurring in the Nernst-Planck equation, is derived. Generally (at sufficiently course spatial resolutions so that the charge density can be assumed to be continuous), $v$ obeys Poisson's equation:
\begin{equation}
\nabla \cdot (\epsilon \nabla v) = -\rho,
\label{poissons}
\end{equation}
where $\epsilon$ is the dielectric constant, and $\rho = \rho_s + F \sum_k (z_k c_k)$ is the total charge density.

In biological tissue, the charge relaxation time $\tau = r_n \epsilon_n$ is very small in any region except in the thin Debye layer ($\sim$1nm) surrounding a bio-membrane. Any nonzero $\rho$ in the bulk solution will decay very rapidly ($\tau\sim$1ns) to zero \cite{Grodzinsky2011}. Several models have simulated electrodiffusion by solving the Nernst-Planck equations in one or more dimensions, with Poisson's equation for $v$  (see e.g., \cite{Leonetti1998a, Lu2007, Lopreore2008, Nanninga2008, Zheng2011}). The advantage with this procedure is that the Poisson-Nernst-Planck (PNP) equations can be implemented in a general way in three-dimensional space. The challenge is then to specify the appropriate boundary conditions for solving Eq.~\ref{poissons} in the vicinity of membranes. Generally, PNP-solvers apply a fine spatial resolution near the membrane, and simulation time steps smaller than the charge-relaxation time \cite{Lopreore2008}. For these reasons, they tend to be extremely computationally demanding \cite{Mori2009}.

The formalism presented in this work belongs to a class of of one-dimensional models, including the cable equation and several electrodiffusive models \cite{Gardner-Medwin1983, Chen2000, Nygren1999, QianNingSejnowski1989, Langlands2009}, which bypasses the computationally heavy PNP-scheme. The physical interpretation of these models is as follows: Any net charge in a volume $A_n\Delta x$ is implicitly assumed to be located in the thin Debye-layer surrounding the capacitive membrane, and is identical to the charge that determines $v_M$. The remainder of the space (i.e., the bulk) will therefore be electroneural ($\rho_{tot} = 0$). Note that any finite volume, enclosing a piece of membrane, will also be electroneutral. This follows from the charge symmetry condition (Eq.~\ref{CS}), constraining the charge on either side of the membrane to be equal in magnitude and opposite in sign. The charge symmetry condition and the electroneutrality condition are in this way closely related. In these electroneutral models, charge relaxation is implicit. This is a plausible assumption at time scales relevant for most biophysical processes. Accordingly, simulations may be run with time-steps ranging from 1 ms to 1 s, depending on the time course of the included membrane mechanisms.

To our knowledge, the formalism summarized in Fig.~\ref{Fbox} is the first biodiffusive model where the intra- and extracellular voltage gradients have been derived from the charge symmetry condition. Eqs.~\ref{VIgrad} and ~\ref{VEgrad} can be interpreted as summarizing all local and global electrical forces driving the system towards electroneutrality.

A natural future ambition would be to generalize the electrodiffusive formalism to 2 or 3 spatial dimensions, so it can address the same 3-dimensional transport problems as PNP-solvers. The challenge will be to formulate the system as a grid of coupled constraints (electroneutrality in the bulk and Eq.\ref{VMold} for $v_M$ across the membrane) for which the Nernst Planck-equations can be solved with time steps much longer than those involved in the charge relaxation process.

\section*{Model and Methods}

\subsubsection*{Astrocytic membrane mechanisms}
The transmembrane ion fluxes in the astrocyte model were:
\begin{eqnarray}
j_{KM} = \frac{g_{K0} f_{Kir}}{F} \left(v_M-e_K\right) - 2P
\label{jKm}
\\
j_{NaM}= \frac{g_{Na0}}{F}\left(v_M-e_{Na}\right) + 3P
\label{jNam}
\\
j_{ClM} = -\frac{g_{Cl0}}{F}\left(v_M-e_{Cl}\right).
\label{jClm}
\end{eqnarray}
Here, $g_{k0}$ are the baseline conductances of the passive K\textsuperscript{+}, Na\textsuperscript{+} and Cl\textsuperscript{-} currents. The currents depend linearly on the difference between $v_M$ and the reversal potential,
\begin{equation}
e_k =  \Psi \log(c_{kE}/{c_{kI}}),
\label{ek}
\end{equation}
for the respective ion types ($k$). The potassium current was modified by the Kir-function \cite{Chen2000}:
\begin{equation}
\begin{split}
f_{Kir}(c_{KE},\Delta v, v_M) = \sqrt{\frac{c_{KE}}{c_{KE0}} } \left[\frac{1+\exp(18.4/42.4)}{1+\exp[(\Delta v+18.5)/42.5]}\right]
\left[\frac{1+\exp[-(118.6+e_{K0})/44.1]}{1+\exp[-(118.6+v_{M})/44.1]}\right]
\end{split}
\label{Kir}
\end{equation}
where $\Delta v = v_m - e_K$, and $e_{K0} (mV)$ is the Nernst potential for potassium at basal concentrations $c_{kE0}$ and $c_{kI0}$.

The K\textsuperscript{+}/Na\textsuperscript{+}-exchanger uses energy (ATPase) to exchange 2 potassium ions with 3 sodium ions. We used a pump-rate per unit area defined by:
\begin{equation}
P(c_{NaI},c_{KE}) = P_{max} \frac{c_{NaI}^{1.5}}{c_{NaI}^{1.5}+K_{NaI}^{1.5}} \frac{c_{KE}}{c_{KE}+K_{KE}}.
\label{Pump}
\end{equation}
\noindent
The maximum pump rate, $P_{max}$, and the threshold concentrations, $K_{NaI}$ and $K_{KE}$, are given in Table 1.

\subsubsection*{Initial conditions}
Initial conditions were determined in the following way: As a starting point, we used $c_{kn0}=c_{knL}$ and $v_{M0}=v_{ML}$ as our initial conditions, where $c_{knL}$ and $v_{ML}$ were the resting concentrations and resting membrane potential found in a previous study \cite{Øyehaug2012}. We then ran a simulation with no system input or output. With the membrane mechanisms included in Eqs.~\ref{jKm} -~\ref{jClm}, the system had a simulated resting state ($c_{knS}$ and $v_{MS}$) which was close to, but not identical to $c_{knL}$ and $v_{ML}$. For all subsequent simulations, we set the initial conditions to the simulated resting conditions ($c_{kn0}=c_{knS}$ and $v_{M0}=v_{MS}$). The estimated values and the values from the literature are given in Table 1.

Prior to all simulations, we defined the static charge densities:
\begin{eqnarray}
\rho_{sI} =  \frac{O_M}{a_I}C_m v_{m0} - F(c_{KI0} + c_{NaI0} - c_{ClI0})
\label{StatQI}
\\
\rho_{sE} = - \frac{O_M}{a_E}C_m v_{m0} -F(c_{KE0} + c_{NaE0} - c_{ClE0}).
\label{StatQE}
\end{eqnarray}
The static charge densities ensure that the total charge density in $I$ and $E$ are consistent with $v_{M0}$, according to Eq.~\ref{rho def}.

\subsubsection*{Comparison of concentrations and charges}
To allow direct comparison with ion concentrations, we represented the charge density in Eq.~\ref{rho def} as an equivalent concentration of unit charges, defined by:
\begin{equation}
c_{en} = c_{Kn}+c_{Nan}-c_{Cln} + \rho_{sn}/F,
\label{ce}
\end{equation}
\noindent
with Eq.~\ref{StatQI} or Eq.~\ref{StatQE} for $\rho_{ns}$. Likewise, we represented the current densities as equivalent unit-charge flux densities, defined by:
\begin{eqnarray}
j_{en}^f = j_{Kn}^f+j_{Nan}^f-j_{Cln}^f
\label{jef}
\\
j_{en}^d = j_{Kn}^d+j_{Nan}^d-j_{Cln}^d
\label{jed}
\end{eqnarray}

\subsubsection*{Implementation}
The model was implemented in Matlab, and the code will be made publicly available at ModelDB (http://senselab.med.yale.edu/modeldb). Simulations were run using the Matlab-solver \emph{pdepe}, which uses variable time steps. For the simulations presented below, we used a maximum time step of 0.1 s, and used 100 segments in the $x$-direction. Improving the resolution had no visible impact on the predicted results. Initial conditions were as listed in Table 1, and the sealed end boundary conditions (Eq.~\ref{bc}) were applied.

\section*{Acknowledgements}
The project was supported by the Research Council of Norway (eVITA program; project numbers 178892 and 178901), and EU Grant 269921 (BrainScaleS). The funders had no role in study design, data collection and analysis, decision to publish, or preparation of the manuscript. We thank John Wyller and Bj{\o}rn F. Nilsen for useful discussions in the initial phase of the project, and Hans Petter Langtangen for useful feedback on the manuscript.

\bibliography{GH_Arxiv}

\end{document}